\documentclass{article}
\usepackage{spconf,amsmath,graphicx}
\usepackage{hyperref}
\usepackage{multirow}
\usepackage{diagbox}
\usepackage{bm}
\usepackage{balance}
\usepackage{color}

\usepackage{makecell}

\title{Audio-Visual Active Speaker Extraction \\ for Sparsely Overlapped Multi-talker Speech}
\name{Junjie Li$^1$, Ruijie Tao$^{3}$, Zexu Pan$^3$, Meng Ge$^{3}$, Shuai Wang$^1$, Haizhou Li$^{1,2,3}$}
\address{$^1$ Shenzhen Research Institute of Big Data, Shenzhen, China \\
$^2$ The Chinese University of Hong Kong, Shenzhen (CUHK-Shenzhen), China \\
$^3$ Department of Electrical and Computer Engineering, National University of Singapore, Singapore }
\begin{document}
\ninept
\maketitle
\begin{abstract}
Target speaker extraction aims to extract the speech of a specific speaker from a multi-talker mixture as specified by an auxiliary reference. Most studies focus on the scenario where the target speech is highly overlapped with the interfering speech. However, this scenario only accounts for a small percentage of real-world conversations. In this paper, we aim at the sparsely overlapped scenarios in which the auxiliary reference needs to perform two tasks simultaneously: detect the activity of the target speaker and disentangle the active speech from any interfering speech. We propose an audio-visual speaker extraction model named ActiveExtract, which leverages speaking activity from audio-visual active speaker detection (ASD). The ASD directly provides the frame-level activity of the target speaker, while its intermediate feature representation is trained to discriminate speech-lip synchronization that could be used for speaker disentanglement. Experimental results show our model outperforms baselines across various overlapping ratios, achieving an average improvement of more than 4 dB in terms of SI-SNR.


\end{abstract}
\section{Introduction}
\label{sec:intro}

At a cocktail party \cite{bronkhorst2015cocktail}, humans possess the remarkable ability to focus their attention on a specific person's speech while ignoring others, also known as selective auditory attention \cite{cherry1953some}. 
To build an automated system equipped with similar capability, Target Speaker Extraction (TSE) is introduced \cite{zmolikova2023neural}.
TSE aims to extract the target speech from a mixture of multiple speakers by utilizing an auxiliary reference, such as a reference speech signal to explore the voice signature cue, or the visual recording to explore the speech-lip synchronization cue.

Previous TSE works,  such as SpeakerBeam \cite{vzmolikova2019speakerbeam}, SpEx \cite{xu2020spex}, and SpEx+ \cite{ge20_interspeech}, primarily focus on instances of highly overlapped speech, which has an overlapping ratio of nearly 100\%. However, in real-world situations, speakers do not continuously talk without interruptions. Instead, they pause intermittently while others speak, resulting in sparsely overlapped speech. For instance,  the overlapping ratio can be less than  20\% in meetings \cite{ccetin2006analysis}.

For the sparsely overlapped speech scenarios, the auxiliary reference in TSE needs to perform two tasks \cite{delcroix2021speaker, hao2021wase}: a) detecting the activity of the target speaker, where the network learns to output a silence signal when the target speaker is inactive and a speech signal when the target speaker is active, and b) disentangling the speech of the target speaker from any interfering speakers when both are active. Based on that, recent studies \cite{delcroix2021speaker, hao2021wase, lin2021sparsely} have proposed to incorporate Voice Activity Detection (VAD) or personalized VAD to extract speaking activity as auxiliary information for guiding speech separation. However, the extraction performance is highly affected by the hard labels (0 or 1) from the VAD output.

\begin{figure}
    \centering
    \includegraphics[width=0.49\textwidth]{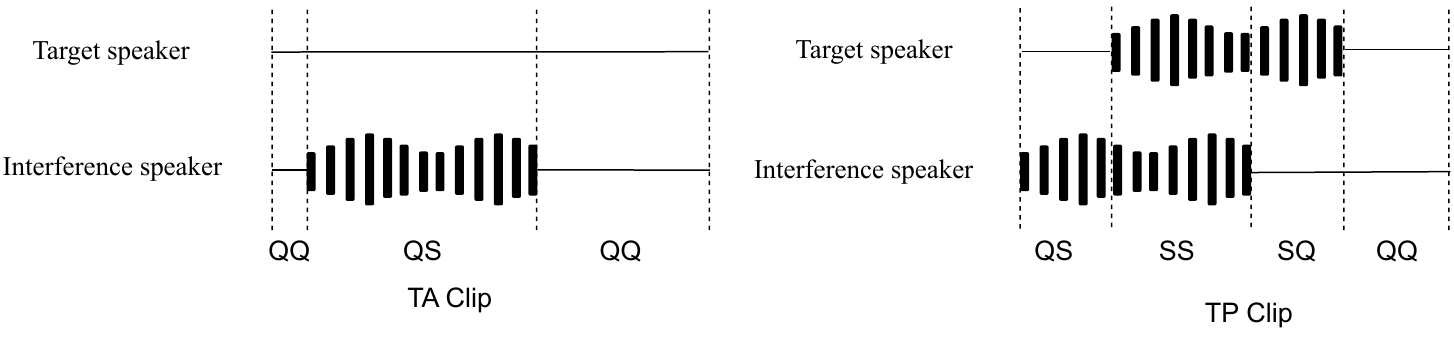}
    \vspace{-5mm}
    \caption{An illustration of real-world speech clips, which can be categorized as Target Absent (TA) and Target Present (TP) at the utterance level. Another finer categorization is according to the status of the target and interference speakers~\cite{pan2022usev} at the segmental level: target Quiet and interference Speaking (QS), target Speaking and interference Speaking (SS), target Speaking and interference Quiet (SQ), target Quiet and interference Quiet (QQ).}
    \label{fig:ta}
    \vspace{-5mm}
\end{figure}

Accounting on the robustness of visual cues against acoustic noise, the use of visual cues in TSE outperforms audio cues for sparsely overlapped speech signals \cite{pan2022usev}. 
 Visual cues, particularly lip movements, provide a high-level discriminative cue for distinguishing speech from non-speech signals for the target speaker \cite{michelsanti2021overview}. Additionally, they offer temporal synchronization information between speech and lip movements \cite{li23ja_interspeech,pan2022selective}.  USEV \cite{pan2022usev} proposes a weighted scenario-aware loss to tackle sparsely overlapped speech with visual cues as the auxiliary reference.    However, it requires knowledge of the start and end times of speech, and manually assigning weights to each scenario loss may result in sub-optimal outcomes. Alternatively, Xiong \cite{xiong2022look} proposes to jointly train ASD and speech enhancement in parallel, yielding improvements in both tasks.  Nevertheless, this approach still struggles to handle the target-absent scenario as the SI-SNR loss used is undefined here.

\begin{figure*}[htbp]
    \centering
    \includegraphics[width=0.9\textwidth]{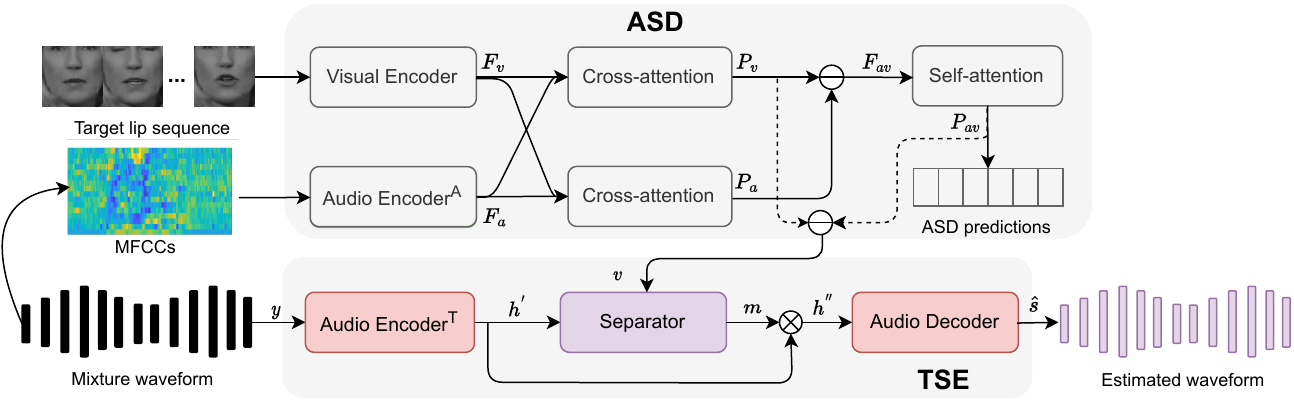}
    \vspace{-5mm}
    \caption{An overview of ActiveExtract, which consists of an ASD module and a TSE module.  $P_v$ denotes attention visual feature and $P_{av}$ denotes speaking activity feature.  $\otimes$ denotes element-wise multiplication and $\ominus$ denotes concatenation over the feature channel dimension.}
    \label{fig:model}
    \vspace{-5mm}
\end{figure*}

While the use of visual cues has proven to be highly effective, there is still room for improvement in sparsely overlapped speech scenarios. This is because prior visual frontends are primarily pretrained from lipreading \cite{wu2019time, afouras2018deep,Afouras18,lin2023av}, which only consider active lip shapes and do not account for inactive speaker cases. Therefore, these pretrained models do not explicitly provide any cues related to speaking activity. In this paper, we present an audio-visual speaker extraction model named ActiveExtract, to utilize audio-visual active speaker detection pretraining instead of lipreading pretraining. The audio-visual ASD pretraining utilizes the speech-lip temporal synchronization to detect the on-screen speaking activity at the frame level, thus, directly providing the activity of the target speaker, while its intermediate feature representation contains the speech-lip synchronization cue that could be used for speaker disentanglement. Thus, the use of the ASD feature eases the efforts in detecting speaker activities, better catering to the sparsely overlapped speech scenarios.

\vspace{-7mm}
\section{Method}
\label{sec:method}

\subsection{Sparsely overlapped speech}

In this paper,  the speech mixture only involves speech signals from two speakers, without any additional noise or reverberation. 
For the TSE task, the target speaker can be either absent or present in a speech mixture. In general, Target Absent (TA) refers to a scenario where the target speaker keeps quiet throughout the conversion, while Target Present (TP) indicates that the target speech is present in the mixture, as shown in Fig \ref{fig:ta}. According to different combination modes of target and interference speech, the speech mixture also can be categorized into four scenarios \cite{pan2022usev, borsdorf2021universal}: QQ, SQ, QS and SS.  QQ denotes a scenario where both target and interference speakers are Quiet. SQ denotes a scenario of Speaking target speaker and Quiet interference speaker. QS denotes Quiet target speaker and Speaking interference speaker. SS denotes both target and interference speakers are Quiet. 

\subsection{Target speaker extraction}

Target speaker extraction aims to extract target speech from a multi-talker speech mixture by utilizing auxiliary information, such as pre-enrolled utterances of the target speaker \cite{subakan2021attention,xu2020spex,ge20_interspeech,vzmolikova2019speakerbeam} or visual frames of the target speaker \cite{lin2023av,pan2022selective,pan2022usev,li23ja_interspeech,wu2019time}. To solve sparsely overlapped speech,  in our approach, we utilize the intermediate features extracted from ASD as the auxiliary information. These features capture the temporal synchronization between lip movements and speech, which can be used to detect speaking activity and extract the target speech.

The TSE module consists of an Audio Encoder, a Separator and an Audio Decoder, as shown in Fig \ref{fig:model}. Audio Encoder employs 1D convolution on mixture waveform $y$ to extract feature of mixture $h'$.  The separator incorporates mixture feature $h'$ and auxiliary reference feature $v$ to model the correlation between different modalities and predict mask $m$. The Audio Decoder then recovers target speech $\hat{s}$ from the element-wise multiplied feature $h^{''}\text{=}m \otimes h'$.  In Section \ref{sec:result}, we will explore the use of  DPRNN \cite{luo2020dual} and AV-sepformer \cite{lin2023av} as the separator backbone for extracting the target speech.

\setlength\tabcolsep{6pt} 
\renewcommand{\arraystretch}{2}
\begin{table*}[htbp]
	\centering
	\fontsize{8}{7}\selectfont
	\caption{A summation of our sparsely overlapped dataset, IEMOCAP-2Mix. }
	\label{tab:data}
	\begin{tabular}{c|c|c|c|c|c|c|c|c|c|c|c|c}
		\hline
		\multirow{2}{*}{Data}& \multirow{2}{*}{Total Clips}& \multirow{2}{*}{TA Clips}&\multicolumn{6}{c|}{TP Clips with different overlapping ratios}& \multicolumn{4}{c}{Duration (hours)}\cr \cline{4-13}

        & & & 0 \% &(0 ,20] \% & (20 ,40] \% & (40 ,60] \% & (60 ,80] \% &(80 ,100] \% & QQ & SQ & QS &SS \cr \hline  
        Training & 20,000 & 758 & 1459 & 1578 & 2597 & 3886 & 4229 & 5493 & 2.06 & 4.72 & 4.68 & 11.98 \cr \hline
        Validation & 5,000 & 158 & 329 & 412 & 634 & 1059 & 1099 & 1309 & 0.55& 1.15 & 1.20 & 2.97 \cr \hline
        Test & 3,000 & 86 & 196 & 233 & 362 & 544 & 712 & 867 & 0.28 & 0.72 & 0.64 & 1.85 \cr \hline

		\hline
	\end{tabular} 
 \vspace{-5mm}
\end{table*}

\subsection{ActiveExtract}

Detecting the speech boundary of the target speaker is a natural approach to address the challenge of sparsely overlapped speech. In ActiveExtract \footnote{Demo: \url{https://activeextract.github.io/}}, we utilize an Audio-Visual ASD module to extract frame-level speaking activity, and the intermediate feature from ASD contains valuable speech-lip synchronization information that can be leveraged to disentangle the target speech.

In our implementation, we adopt TalkNet \cite{tao2021someone} as our ASD module. TalkNet consists of a feature representation frontend, which includes a Visual Encoder and an Audio Encoder, and a speaker detection backend classifier comprising two Cross-attention networks and a Self-attention network. The Visual Encoder captures spatial information within each visual frame, aiming to learn the representation of lip dynamics denoted as $F_v$. On the other hand, the Audio Encoder learns an audio content representation denoted as $F_a$ from a vector of Mel-frequency cepstral coefficients (MFCCs). The Cross-attention modules in TalkNet model the interaction between audio and video features to generate aligned audio and visual features. For instance, they produce attention audio feature $P_a$ and attention visual feature $P_v$. Following the Cross-attention modules, a Self-attention module is applied to learn an audio-visual temporal speaking activity feature denoted as $P_{av}$, which helps distinguish between speaking and non-speaking frames.

$P_v$ represents the lip motion and can be utilized to disentangle the target speech. $P_{av}$ is a correlation representation between audio and visual features, which can be employed to discriminate between speech and non-speech signals. The dimensions of $P_v$ and $P_{av}$ are 128 and 256, respectively, and their lengths correspond to the number of visual frames.
We concatenate these features along the channel dimension to obtain a new feature $v$, which serves as auxiliary information to guide the target speaker extraction process.

\subsection{Loss function}
To better handle the SS scenario, we first pretrain the ActivExtract model on highly overlapped speech, then finetune it on the sparsely overlapped speech.

For pretraining, we adopt Signal-to-Distortion Ratio (SDR) \cite{le2019sdr} as the loss function. Since there are no TA scenarios in the pretraining dataset, the SDR loss is suitable: 
\begin{equation}
    \mathcal{L}_{\text{SDR}} = -10\log_{10}\frac{||s||^2}{||\hat{s}-s||}
    \label{equ:sdr}
\end{equation}
where $s$ is the ground truth and $\hat{s}$ is estimated speech. 

For finetuning,  Source-Aggregated SDR (SA-SDR) \cite{von2022sa} is adopted as the loss function. SA-SDR concatenates all output channels to compute a global SDR, which is able to manage TA clips:
\begin{align}
    \mathcal{L}_{\text{SA-SDR}} = -10\log_{10}\frac{\sum\limits_{k=1}^K||s_k||^2}{\sum\limits_{k=1}^{K}||s_k-\hat{s_k}||^2}
    \label{equ:sasdr}
\end{align}
where $K$ denotes the number of utterances in a training batch in this paper.

For finetuning,  we also adopt Scenario-Aware Differentiated Loss (SADL) \cite{pan2022usev} for comparison. SADL applies different loss functions based on the scenario. For SS and SQ segments where the target speaker is active, it applies $\mathcal{L}_{\text{SDR}}$ as the loss function. For QQ and QS segments where the target speaker is inactive, it applies the energy of the segments $\mathcal{L}_\text{E}$ as the loss function. Then the total loss $\mathcal{L}_{\text{SADL}}$ is a weighted sum of the 4 differentiated loss values as follows:
\begin{equation}
    \mathcal{L}_{\text{SADL}} =\alpha\mathcal{L}_{\text{E}}^{QQ}+\beta \mathcal{L}_{\text{SDR}}^{SQ} +\gamma \mathcal{L}_{\text{SDR}}^{SS} +\delta \mathcal{L}_{\text{E}}^{QS}
    \label{equ:sadl}
\end{equation}
where
\begin{equation}
    \mathcal{L}_{\text{E}} = 10\log_{10}(||\hat{s}||)
\end{equation}
and $\alpha$, $\beta$, $\gamma$ and  $\delta$  are  weights for the loss. Compared to $\mathcal{L}_{\text{SA-SDR}}$, $\mathcal{L}_{\text{SADL}}$ requires the start and end times of target speech and interference speech, and needs to manually assign weights to each loss. 


\section{Experimental setup }
\label{sec:exp}

\subsection{Dataset}

1) \textbf{Highly overlapped speech dataset}: VoxCeleb2-2Mix, which is used for pretraining.  VoxCeleb2 \cite{chung2018voxceleb2} contains 1 million synchronized videos from over 6,112 celebrities. To construct a highly overlapped speech mixture, we randomly select two audio utterances from different speakers and mix them at various Signal-to-Noise Ratio (SNR) ranging from -10 dB to 10 dB. Each utterance is trimmed to a duration of 4 seconds.  Audios are sampled at 16 kHz, and the videos are sampled at 25 FPS.  We randomly create 5,000 speech mixture clips from the VoxCeleb2 training set as our validation set. Note that our training set is dynamically mixed during the training process, allowing for a diverse range of mixture configurations.

\setlength\tabcolsep{4pt} 
\renewcommand{\arraystretch}{2}
\begin{table*}[htbp]
	\centering
	\fontsize{8}{6.8}\selectfont
	\caption{Performance on the test set of IEMOCAP-2Mix with different overlapping ratios. $\downarrow$ denotes that a lower value indicates better performance. $\uparrow$ denotes that a higher value indicates better performance. $\mathcal{L}_{\text{SADL}}^o$ and  $\mathcal{L}_{\text{SADL}}^b$ adopt different weight settings of each scenario loss.  `*' denotes we replicate the model and retrain it on our dataset. We have also chosen different model structures as separator for comparison. }
	\label{tab:result}
	\begin{tabular}{c|c|c|c|c|c|c|c|c|c|c|c}
		\hline
		 \multicolumn{2}{c|}{\multirow{2}{*}{Model}} &\multicolumn{2}{c|}{Configurations} &\multirow{2}{*}{TA Power $\downarrow$}&\multicolumn{7}{c}{TP  SI-SNR  $\uparrow$}\cr \cline{6-12} \cline{3-4}

        \multicolumn{2}{c|}{} &Separator &Loss & & 0 \% &(0 ,20] \% & (20 ,40] \% & (40 ,60] \% & (60 ,80] \% &(80 ,100] \% &Avg.  \cr \hline  \hline
        
        \multicolumn{2}{c|}{Mixture}  & - & -  & 18.62 & 63.57 & 0.38 & 0.44 & -0.13 & -0.24 & -0.12 & 4.24  \cr \hline

        \multicolumn{2}{c|}{\multirow{3}{*}{USEV$^*$ \cite{pan2022usev}}}  &\multirow{3}{*}{DPRNN}& $\mathcal{L}_{\text{SADL}}^o$&18.61 & 40.59 & 0.37 & 0.49 & -0.16 & -0.20 & -0.13 & 2.70  \cr 
         \multicolumn{2}{c|}{} & & $\mathcal{L}_{\text{SADL}}^b$ & -2.91 & 29.13& 9.44& 6.87& 3.44& 2.61& 1.56& 5.31\cr 

         \multicolumn{2}{c|}{}  & & $\mathcal{L}_{\text{SA-SDR}}$ &-15.38 &17.37  &8.06 &6.40 &3.62 &2.69  &1.66 &4.43 \cr \hline

        \multicolumn{2}{c|}{\multirow{2}{*}{AV-Sepformer$^*$ \cite{lin2023av}} } &\multirow{3}{*}{AV-Sepformer} & $\mathcal{L}_{\text{SA-SDR}}$ &     -63.22 &19.50 & 5.47 & 5.07 & 3.12 & 2.28 & 1.76 & 4.03  \cr 
        \multicolumn{2}{c|}{} & & $\mathcal{L}_{\text{SADL}}^b$&-61.89  &25.10  &7.88 &6.88 &4.82&3.92&3.84 &6.17  \cr 
        \cline{1-2} \cline{4-12}

        \multicolumn{2}{c|}{ASD+AV-Sepformer}  & & $\mathcal{L}_{\text{SDR}}$ &  -54.01 & 16.60 & 7.52 & 6.51 & 4.96 & 3.36 & 3.35 & 5.27     \cr  \hline
        
        \multirow{5}{*}{\textbf{ActiveExtract}}& \multirow{3}{*}{$P_{av}\& P_v$}  &DPRNN  & $\mathcal{L}_{\text{SA-SDR}}$&-20.04& 20.32 &9.15  &7.73 &5.20 &4.00 &3.86 &6.15   \cr

         & & \multirow{2}{*}{AV-Sepformer} &$\mathcal{L}_{\text{SA-SDR}}$ & \textbf{-64.69} & 35.14 & 9.76& 10.00 &7.18& 6.30& 6.68& 9.25 \cr  


 & &&$\mathcal{L}_{\text{SADL}}^b$ &-49.41&\textbf{43.14}  &
        11.18  &10.25 & 8.32 &7.33 &\textbf{7.58} &\textbf{10.67}   \cr \cline{2-12}
  &$P_v$ & \multirow{2}{*}{AV-Sepformer} & \multirow{2}{*}{$\mathcal{L}_{\text{SADL}}^b$} & -57.57 & 34.47& \textbf{11.89}& 10.39& \textbf{8.47}& \textbf{7.41}& 7.34&10.13 \cr 
   &$P_{av}$ & & &-47.05 &39.07 &11.85 &\textbf{10.42} &8.16 &7.16 &6.91&10.20\cr



		\hline
	\end{tabular} 
	\vspace{-5mm}
\end{table*}

2) \textbf{Sparsely overlapped speech dataset}: IEMOCAP-2Mix is used to finetune and test our proposed model. We create sparsely overlapped mixtures by selecting 2 utterances from the Interactive Emotional Dyadic Motion Capture (IEMOCAP) dataset \cite{busso2008iemocap}. IEMOCAP consists of 12 hours of 150 dyadic multi-modal conversations, each containing 2 speakers, with a total of 10 speakers in the dataset. It covers speaking and quiet scenarios in conversations. We follow the setting \footnote{\url{https://github.com/zexupan/USEV}} in previous work \cite{pan2022usev} to create the IEMOCAP-2Mix dataset. Audios 
are sampled at 16 kHz and videos are sampled at 25 FPS. And there are  20,000, 5,000 and 3,000 utterances in training, validation and test set, respectively\footnote{The amount of training set used in \cite{pan2022usev} is 20 times larger than ours.}. The duration of the utterances in our dataset ranges from 3 to 6 seconds. The mixing SNR is set between -10 dB to 10 dB. Due to the limited number of speakers in IEMOCAP, speakers in the above three sets are overlapped but utterances are not.
IEMOCAP-2Mix encompasses speech under various scenarios, resulting in different overlapping ratios, as reported in Table \ref{tab:data}. It contains TA clips and TP clips, and the overlapping ratio is calculated by dividing the duration of the overlapped segment by the total duration of no-silence speech, i.e., SS/(SQ+QS+SS).

\subsection{Implementation details}

\subsubsection{Training procedure}
\label{sec:detail}

The overall training of ActiveExtract consists of three stages\footnote{Code: \url{https://github.com/mrjunjieli/ActiveExtract}}.

1) ASD module is first pretrained using TalkSet \cite{tao2021someone}, which is selected from LRS3 \cite{afouras2018lrs3} and VoxCeleb2 \cite{chung2018voxceleb2}. 

2) To better handle overlapped speech scenarios,  the entire ActiveExtract is first pretrained on highly overlapped speech dataset, VoxCeleb2-2Mix, using $\mathcal{L}_{\text{SDR}}$.  

3) To adapt to sparsely overlapped speech scenarios, the entire ActiveExtract is finetuned on sparsely overlapped speech dataset, IEMOCAP-2Mix, using $\mathcal{L}_{\text{SA-SDR}}$ or $\mathcal{L}_{\text{SADL}}$.

\subsubsection{Training setup}

For the pretraining and finetuning stage, we set the initial learning rate as 10e-4 with Adam optimizer. 
If the validation loss has not been decreased for 3 epochs, we halve the learning rate. If the best validation loss has not been improved for 10 consecutive epochs, we stop the training procedure. 
We set the maximum number of training epochs to 100 and 30 for the pretraining stage and finetuning stage, respectively.

\subsubsection{Baselines}

USEV \cite{pan2022usev} consists of a lipreading pretrained visual frontend and a DPRNN \cite{luo2020dual} separator. The L, B, N, R and K in USEV are set to 40, 64, 256, 6 and 100 according to \cite{pan2022usev}.

AV-Sepformer \cite{lin2023av} also has a lipreading pretrained visual frontend. Its separator consists of  IntraTransformer, CrossModalTransformer and InterTransformer. Different from \cite{lin2023av}, we set  L, C, $\text{N}_{\text{intra}}$, $\text{N}_{\text{inter}}$, N and $\text{N}_{\text{head}}$ to 16, 160, 4, 4, 256 and 8. 

ASD+AV-Sepformer refers to a sequential combination of ASD and AV-Sepformer. In this case, the AV-Sepformer is finetuned on TP clips from IEMOCAP-2Mix using $\mathcal{L}_{\text{SDR}}$.  
If ASD predicts the target speaker as active, then this system outputs speech estimated from AV-Sepformer. Otherwise, the system outputs zero. 



\subsection{Evaluation metrics}

To evaluate the performance of recovering target signal and suppressing interference speech, we utilize two metrics: SI-SNR (dB) \cite{le2019sdr,luo2019conv} and Power (dB/s) \cite{pan2022usev}. These metrics are shown as follows: 

\begin{equation}
    \text{SI-SNR} = 10 \log_{10}\frac{||\frac{<\hat{s},s>s}{||s||^2}||^2}{||\hat{s}-\frac{<\hat{s},s>}{||s||^2}||}
\end{equation}

\begin{equation}
    \text{Power}=10 \log_{10}\frac{||\hat{s}||^2}{T_{s}}
\end{equation}
where $T_s$ denotes duration of 
$\hat{s}$ in seconds. The SI-SNR metric is used to evaluate the performance when the target signal is present, while the Power metric assesses the performance when the target signal is absent.

\section{Results}
\label{sec:result}

\subsection{Comparative studies}
Table \ref{tab:result} presents a comparison of model performance using different separators and loss functions across various overlapping ratios. The weights  $\{\alpha, \beta, \gamma, \delta \}$ used in  $\mathcal{L}_{\text{SADL}}^o$ and $\mathcal{L}_{\text{SADL}}^b$ are set to $\{0.005,1,1,0.005\}$ and $\{0.0005,0.1,1,0.005\}$, respectively. 
$\mathcal{L}_{\text{SADL}}^o$ represents the original weight settings from USEV \cite{pan2022usev}, while $\mathcal{L}_{\text{SADL}}^b$ represents the best weight settings discovered in our experiments.

1) Loss function:  
Irrespective of the model used, $\mathcal{L}_{\text{SADL}}$
  consistently outperforms $\mathcal{L}_{\text{SA-SDR}}$ on TP clips, whereas 
$\mathcal{L}_{\text{SA-SDR}}$ exhibits superior performance on TA clips.
This difference in performance could be attributed to the weights assigned to the SQ and SS scenarios in $\mathcal{L}_{\text{SADL}}$. It is worth noting that the performance of 
$\mathcal{L}_{\text{SADL}}$ is highly dependent on the weights assigned to each scenario, resulting in a substantial performance gap between 
$\mathcal{L}_{\text{SADL}}^o$ and $\mathcal{L}_{\text{SADL}}^b$. However, regardless of the loss function used, ActiveExtract demonstrates improved performance compared to the baselines by incorporating speaking activity information through the ASD module.

2) Separator: 
Then we compare the performance of ActiveExtract using different separator backbones, such as AV-Sepformer separator and DPRNN separator. For both separator backbones, ActiveExtract demonstrates significant improvements over their respective baselines. Particularly, when AV-Sepformer was employed as the separator, ActiveExtract demonstrated an average improvement of more than 4 dB compared to AV-Sepformer baseline on TP clips. 
This finding indicates that our ASD pretrained visual frontend is more suitable than a lipreading pretrained visual frontend for real-world conversations. However, the performance of ASD+AV-Sepformer is relatively poor, as it heavily relies on the accuracy of the binary output from ASD.  Any incorrect predictions by ASD significantly impact the system's performance. That is why we utilize intermediate features of ASD rather than its binary output to design our model ActiveExtract.

\subsection{Ablation studies}
When considering both $P_v$ and $P_{av}$, the model achieves a significant improvement than previous methods. When considering $P_{av}$ or $P_v$ only, the model still achieves competitive performance, although slightly worse than considering both. This suggests that both $P_v$ and $P_{av}$ features contribute to the overall performance and complement  each other. 
Moreover, when considering $P_{av}$ only, system can perform slightly better than using $P_{v}$ only across different overlapping ratios, except 0\% overlapping ratio scenario. This finding suggests that $P_{av}$
  contains more valuable speaking activity information and proves to be more effective in the task of target speaker extraction. On the other hand, $P_v$  plays a crucial role to avoid speech distortion.


\section{Conclusions}

In this paper, we propose ActiveExtract, an innovative audio-visual model specifically designed to tackle the challenge of sparsely overlapped speech. By leveraging speaking activity information, ActiveExtract effectively distinguishes between speech and non-speech signals, enabling accurate extraction of the target speaker. Our experimental results demonstrate the remarkable performance of ActiveExtract in extracting the target speech while suppressing interference from other speakers, which effectively addresses the problem of target speaker extraction on real-world conversations.

\clearpage
\bibliographystyle{IEEEbib}
\bibliography{refs}
\end{document}